# Magnetic properties of the spin-1 two-dimensional $J_1$-$J_3$ Heisenberg model on a triangular lattice

P. RUBIN[a,*], A. SHERMAN[a], M. SCHREIBER[b]

[a] Institute of Physics, University of Tartu, Riia 142, 51014 Tartu, Estonia

[b] Institut für Physik, Technische Universität, D-09107 Chemnitz, Germany

Motivated by the recent experiment in $NiGa_2S_4$, the spin-1 Heisenberg model on a triangular lattice with the ferromagnetic nearest- and antiferromagnetic third-nearest-neighbor exchange interactions, $J_1 = -(1-p)J$ and $J_3 = pJ$, $J > 0$, is studied in the range of the parameter $0 \le p \le 1$. Mori's projection operator technique is used as a method, which retains the rotation symmetry of spin components and does not anticipate any magnetic ordering. For zero temperature several phase transitions are observed. At $p \approx 0.2$ the ground state is transformed from the ferromagnetic order into a disordered state, which in its turn is changed to an antiferromagnetic long-range ordered state with the incommensurate ordering vector $\mathbf{Q}' \approx (1.16, 0)$ at $p \approx 0.31$. With growing $p$ the ordering vector moves along the line $\mathbf{Q}' - \mathbf{Q}_c$ to the commensurate point $\mathbf{Q}_c = (2\pi/3, 0)$, which is reached at $p = 1$. The final state with the antiferromagnetic long-range order can be conceived as four interpenetrating sublattices with the 120° spin structure on each of them. Obtained results offer a satisfactory explanation for the experimental data in $NiGa_2S_4$.

Investigation of the spin-1 Heisenberg model on a two-dimensional (2D) triangular lattice is of interest in connection with recently synthesized compound $NiGa_2S_4$ [1]. This crystal is characterized by a spin disorder at low temperatures and strong antiferromagnetic correlations. Magnetism in this system is mainly connected with $Ni^{2+}$ ($S$=1) ions, which form a 2D triangular lattice. Experiments on neutron scattering reveal a short-range order, being characterized by an incommensurate ordering vector. For the description of the observed order the classical $J_1$-$J_3$ 2D Heisenberg model on a triangular lattice was proposed [1]. The dominating antiferromagnetic interaction $J_3$ between the third nearest neighbors (TNN) and the weak ferromagnetic interaction $J_1$ between the nearest neighbors (NN) were taken into account. This model agrees with first-principle calculations of the exchange parameters [2].

In this work the corresponding quantum model with $S$=1 is considered. The Hamiltonian of the model reads

$$H = \frac{1}{2}\sum_{\mathbf{nm}} J_{\mathbf{nm}} \left( s_{\mathbf{n}}^z s_{\mathbf{m}}^z + s_{\mathbf{n}}^+ s_{\mathbf{m}}^- \right), \qquad (1)$$

where $s_{\mathbf{n}}^z$ and $s_{\mathbf{n}}^\pm$ are components of the spin-1 operators, $\mathbf{n}$ and $\mathbf{m}$ label sites of the triangular lattice. Only the NN and TNN interactions are taken into account,

$$J_{\mathbf{nm}} = \sum_{\mathbf{a}} \left( J_1 \delta_{\mathbf{n},\mathbf{m}+\mathbf{a}} + J_3 \delta_{\mathbf{n},\mathbf{m}+2\mathbf{a}} \right),$$

with the vectors $\mathbf{a}$ and $2\mathbf{a}$ connecting the NN and TNN sites. We use the lattice spacing $a = |\mathbf{a}|$ as the unit of length. The exchange interactions, $J_1 = -(1-p)J$ and $J_3 = pJ$, $J > 0$, are expressed through the parameter of frustration $p$, which varies in the range from 0 to 1 ($J$ is set as the unit of energy). Mori's projection operator technique [3] gives the following expression for the spin Green's function

$$D(\mathbf{k}\omega) = \frac{6J\left[(1-p)C_1(1-\gamma_{\mathbf{k}}) - pC_{2a}(1-\gamma_{2\mathbf{k}})\right]}{\omega^2 - \omega_{\mathbf{k}}^2}, \qquad (2)$$

where $\gamma_{\mathbf{k}} = \frac{1}{3}\cos(k_x) + \frac{2}{3}\cos(k_x/2)\cos\left(k_y\sqrt{3}/2\right)$, the frequency of spin excitations $\omega_{\mathbf{k}}$ contains products of $1-\gamma_{\mathbf{k}}$, $1-\gamma_{2\mathbf{k}}$ and spin correlations such as $C_1 = \left\langle s_{\mathbf{n}}^+ s_{\mathbf{n}+\mathbf{a}}^- \right\rangle$ and $C_{2a} = \left\langle s_{\mathbf{n}}^+ s_{\mathbf{n}+2\mathbf{a}}^- \right\rangle$. These correlations can be expressed through $D(\mathbf{k}\omega)$ and calculated self-consistently. Details of our approach can be found in [4].

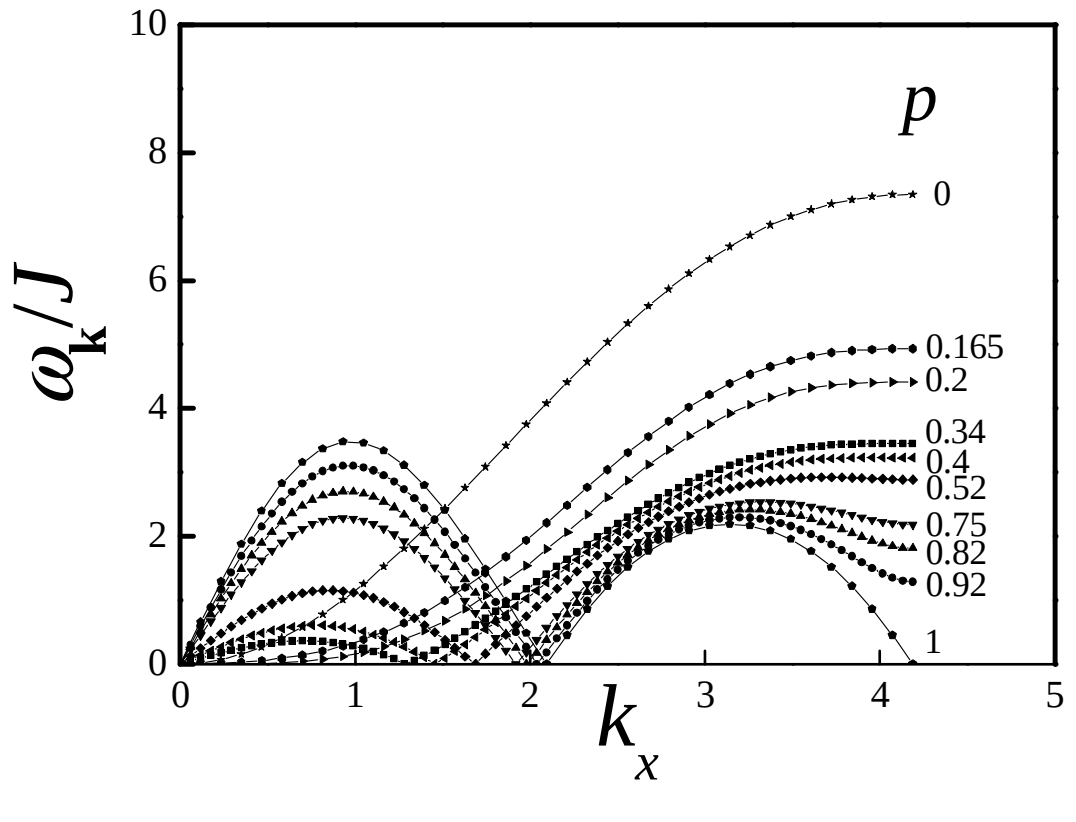

Fig.1. The dispersion of spin excitations $\omega_{\mathbf{k}}$ (in units of $J$) along the $X$ axis for different values of the frustration parameter $p$ at $T$=0 [5].

With changing $p$ the ground state (GS) of the model undergoes a number of transformations. The spectra of magnetic excitations at different values of $p$ and $T$=0 are shown in Fig. 1. In the range $0 \le p \le 0.2$ the GS of the system is ferromagnetic: near the Γ-point the dispersion $\omega_{\mathbf{k}}$ is parabolic, which is typical for a ferromagnet. With the growth of the antiferromagnetic exchange $J_3$ the system passes into a disordered state. When the value $p \approx 0.31$ is reached the system goes into the state with the long-range antiferromagnetic order with the incommensurate ordering vector $\mathbf{Q}' \approx (1.16, 0)$. The frequency of magnetic excitations vanishes at this wave vector (see Fig. 1). With growing the frustration parameter $p$ the ordering vector moves along the $X$ axis from the point $\mathbf{Q}'$ to $\mathbf{Q}_c = (2\pi/3,0)$, which is reached at $p$=1. For this $p$ the incommensurate antiferromagnetic order transforms to a commensurate phase, which can be conceived as four interpenetrating 120° spin structures on sublattices with twice as large lattice spacing. Spin orientations in these sublattices are independent of one another.

*corresponding author; e-mail: rubin@fi.tartu.ee

At $p\approx0.82$ the calculated ordering vector is close to that observed in $NiGa_2S_4$ [1]. For the finite temperature of the experiment, in accord with the Mermin-Wagner theorem, our long-range order passes into the short-range one. However, its calculated correlation length is an order of magnitude larger than the experimental value.

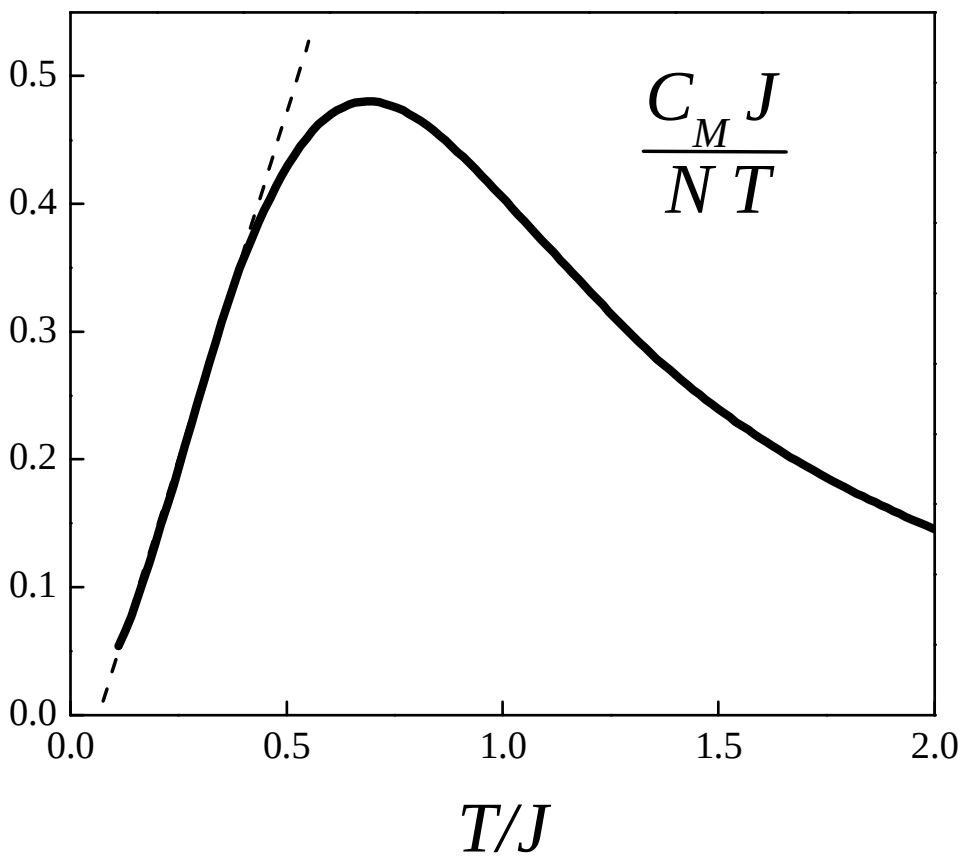

Fig.2. The temperature dependence of the specific heat (the solid line) obtained in the $J_1$-$J_3$ model. The dashed line demonstrates the linear variation of $C_M/T$ with $T$. $N$ is the number of sites.

In $NiGa_2S_4$, the magnetic part of the specific heat exhibits quadratic temperature behavior [1]. As seen in Fig. 2, the $J_1$-$J_3$ Heisenberg model is able to reproduce this dependence, which extends up to $T/J \approx 0.4$. Such temperature behavior is related to the quasilinear spin-excitation dispersion near its minima. The calculated uniform spin susceptibility shown in Fig. 3 also has the same shape as the susceptibility in $NiGa_2S_4$.

In summary, Mori's projection operator technique was used for investigating magnetic properties of the quantum $J_1$-$J_3$ Heisenberg model on a triangular lattice. At zero temperature, depending on the ratio $J_1/J_3$ between the ferromagnetic nearest and the antiferromagnetic third nearest neighbor interactions, the system is characterized by the ferromagnetic ordering, spin disorder, incommensurate and commensurate antiferromagnetic ordering. At $J_1/J_3 \approx -0.22$ the model describes key features observed [1] in $NiGa_2S_4$ – the incommensurate short-range antiferromagnetic order at finite temperature, the quadratic temperature dependence of specific heat and the shape of the uniform susceptibility.

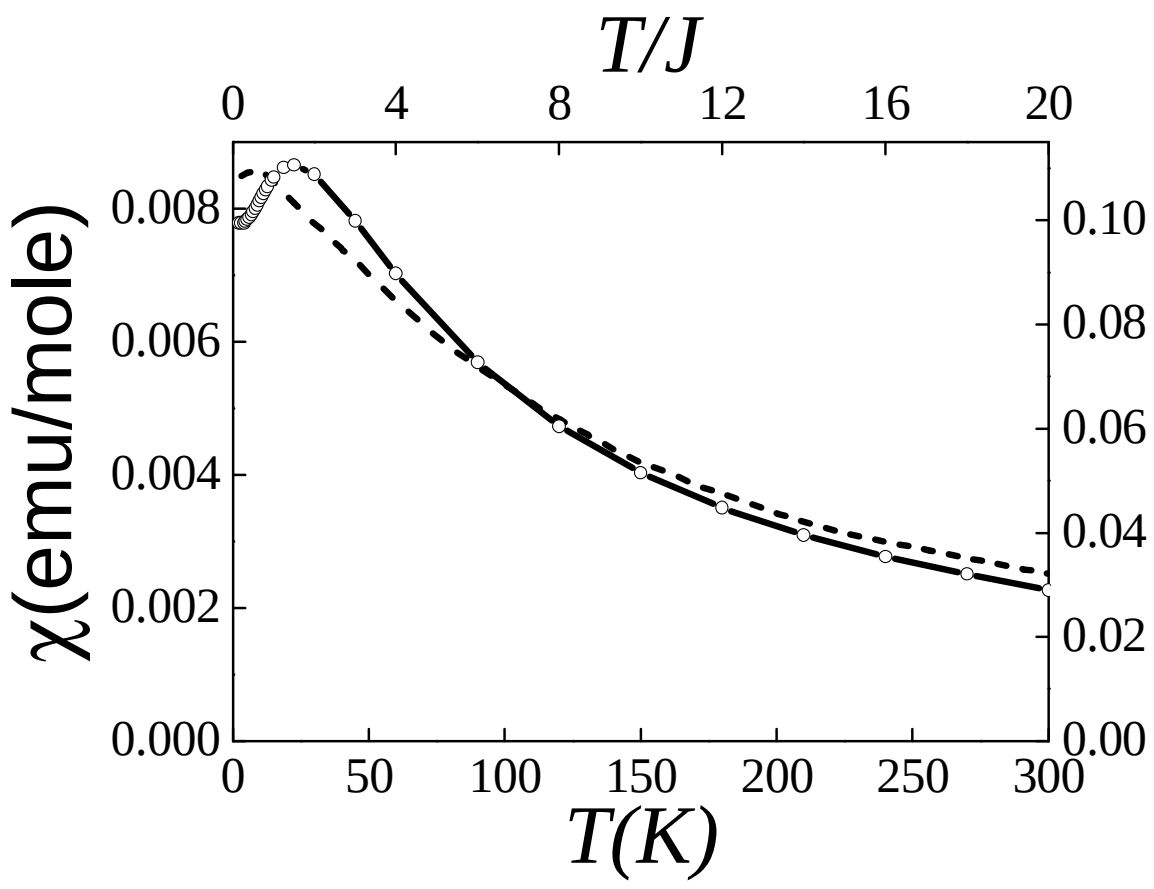

Fig.3. The temperature dependences of the uniform magnetic susceptibility χ in $NiGa_2S_4$ [1] (the dashed line, the left axis) and in the model (the solid line, the right axis).

**Acknowledgement**

This work was supported by the European Regional Development Fund (Centre of Excellence "Mesosystems: Theory and Applications", TK114) and the ESF Grant No. 9371.